\documentclass[a4paper]{article}

\usepackage{Clarity2023}

\usepackage{xcolor}
\usepackage{hyperref}

\newcommand{\gt}{\ensuremath >}

\definecolor{darkred}{rgb}{0.8, 0.0, 0.0}

\title{TF-MLPNet: Tiny Real-Time Neural Speech Separation}

\name{Malek Itani,  Tuochao Chen, Shyamnath Gollakota}
\address{Paul G. Allen School of Computer Science \& Engineering, University of Washington}
\email{\{malek,tuochao,gshyam\}@cs.washington.edu}

\newcommand{\xref}[1]{\S\ref{#1}}
\begin{document}

\maketitle
\begin{abstract}

Speech separation on hearable devices can enable transformative augmented and enhanced hearing capabilities. However, state-of-the-art speech separation networks cannot run in real-time on tiny, low-power neural accelerators designed for  hearables, due to their limited compute capabilities. We present {\it TF-MLPNet,} the first speech separation network capable of running in real-time on such low-power accelerators while outperforming  existing streaming models for  blind speech separation and target speech extraction. Our network operates in the time-frequency domain, processing frequency sequences with stacks of fully connected layers that alternate along the channel and frequency dimensions, and independently processing the time sequence at each frequency bin using convolutional layers. Results show that our mixed-precision quantization-aware trained (QAT) model  can process 6 ms audio chunks in real-time on the GAP9 processor, achieving  a 3.5-4x  runtime reduction compared to prior speech separation models.
 
\end{abstract}

%
\section{Introduction}
\label{sec:intro}

Over the past decade, two key technological trends have emerged. First, deep learning has become central to  speech separation algorithms~\cite{tse,tfgridnet,sepformer,dprnn}, which typically require large, energy intensive resources like GPUs. Second, there is increasing interest in incorporating speech separation into hearables, such as hearing aids, headphones, and earbuds, to develop advanced augmented and enhanced hearing applications that program acoustic scenes and address the cocktail party problem~\cite{semantichearing,lookoncetohear,cornell2023multichanneltargetspeakerextraction,soundbubble}. However, these small, power-constrained devices have limited computing capabilities. To bridge this gap, several hardware efforts have focused on developing tiny, low-power neural network accelerators~\cite{syntiant,greenwaves}.  These platforms are, however, significantly more constrained than both GPUs and  general-purpose embedded CPUs like Raspberry Pi. 



State-of-the-art speech separation networks cannot run in real-time on  low-power neural accelerators for hearables. These networks operate in the time-frequency (TF) domain, modeling time and frequency components as sequences using recurrent networks~\cite{fspen,semamba}, attention mechanisms~\cite{mpsenet}, or both~\cite{tfgridnet}. At each time interval, the 
frequency sequence sub-network processes the entire sequence of frequencies, while the time sequence sub-network independently processes across time, at each of the frequency bins. Streaming applications like enhanced hearing however  require neural networks to operate in real-time on small blocks ($\le$10 ms), which imposes significant  computational and algorithmic constraints.

\begin{figure}[t!]
\centering
\vskip -0.15in
\includegraphics[width=\linewidth]{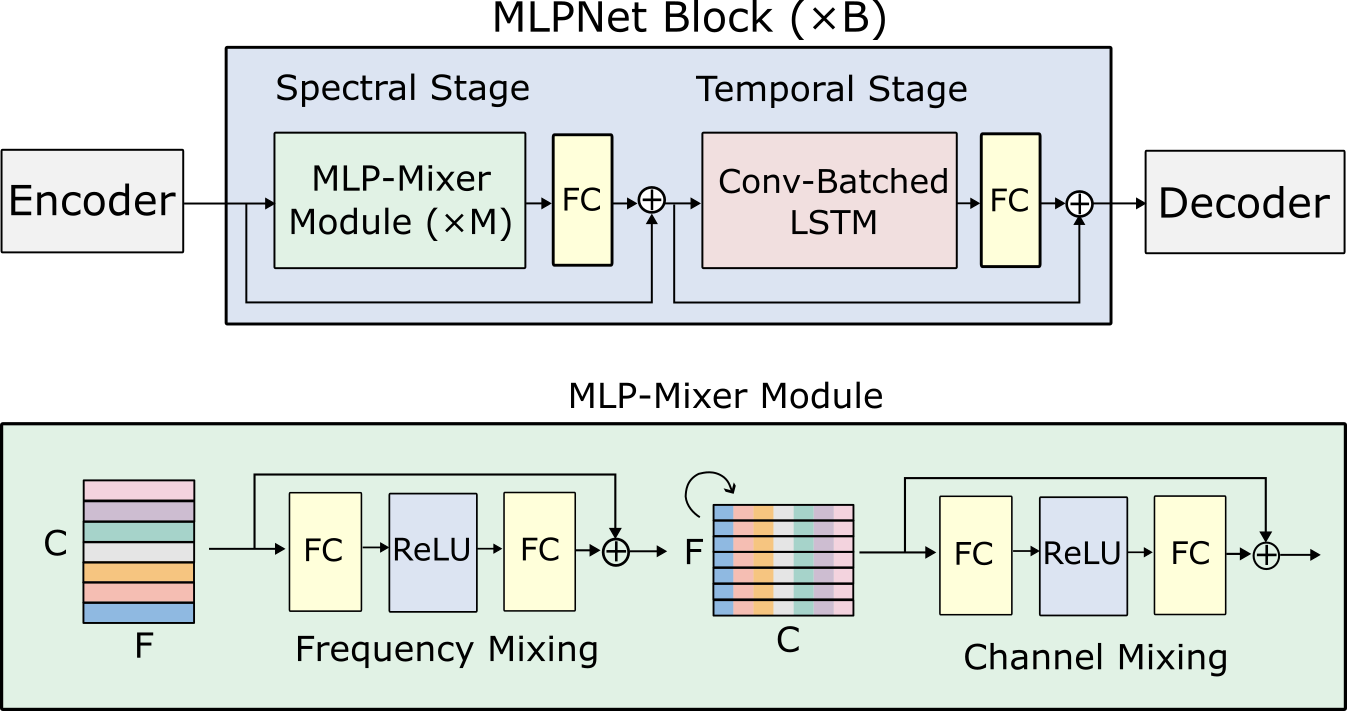}
\vskip -0.1in
\caption{{The TF-MLPNet architecture has  a conv-batched LSTM in the temporal stage, enabling parallel inference of batched LSTM inputs via convolutional layers, and   a highly parallel all-MLP-Mixer module in the spectral stage, replacing the sequential bidirectional LSTM. (FC: fully connected layer)}}
\label{fig:architecture}
\vskip -0.2in
\end{figure}

We introduce {\it TF-MLPNet}, the first on-device real-time neural speech separation network for low-power hearables. We make two key observations: 1) Using recurrent models to process frequency sequences sequentially slows computation, while transformer and state-space architectures are incompatible with our target low-power accelerators; 2) Typically, time sequence modeling at each frequency bin are parallelized using batched processing. However, our target  accelerators are designed for low-power inference and  operate on one input at a time~\cite{syntiant, greenwaves}, i.e., with a batch size of 1. This precludes batched processing, leading again to slow, sequential computation.

Inspired by all-MLP architectures~\cite{mlpmixer,hyperconformer,summarymixing}, our  model (Fig.~\ref{fig:architecture}) replaces the recurrent networks used for  processing frequency bins, with stacks of fully connected layers applied alternately along the channel and frequency dimensions, enabling parallel processing across frequency bins for improved efficiency. We also introduce a hardware-software co-design method that parallelizes time sequence processing at each frequency bin. By using just convolutional layers with a batch size of 1, we effectively parallelize batched LSTMs, mimicking the batched inference of a recurrent network in a single time step. Finally, we also present  a mixed-precision QAT strategy that balances performance and runtime. Since different modules in the network have varying sensitivity to quantization and its impact on runtime, we apply distinct  quantization configurations to different sub-components in our network,  optimizing performance while maintaining real-time constraints.

We evaluate this architecture on two tasks 1) two-speaker blind speech separation~(BSS) and 2) target speech extraction~(TSE). We  compare TF-MLPNet with multiple variants of the causal  TF-GridNet model~\cite{tfgridnet} for the BSS task and with  pDCCRN~\cite{pdccrn}, and TinyDenoiser~\cite{tinydenoiser} for the TSE task. Our results show that 
 TF-MLPNet achieves state-of-the-art real-time on-device performance. Furthermore, our real-time mixed-precision quantized model results in a  performance drop of only 0.6 dB, compared to a fully floating-point network.




\section{Related work}
{\bf Blind speech separation and target speaker extraction.} Prior neural architectures~\cite{convtasnet, tfgridnet, mossformer, sepreformer} use components like convolutional~\cite{convtasnet}, LSTM~\cite{dprnn}, transformer~\cite{sepformer}, and state-space~\cite{spmamba} layers. However, these models prioritize speech quality over real-time, on-device, or low-power constraints. Furthermore, transformers and state-space models have runtimes and memory demands that exceed our target hardware's capabilities. 


\vskip 0.02in\noindent {\bf Low-latency speech processing.} For  augmented hearing, minimizing input-to-output latency is crucial but can reduce performance due to the limited information available for predicting  output~\cite{lowlatse}. Prior work has explored  architectures for low-latency speech tasks~\cite{percepnet,speakerbeamss,deepfilternet2,lookoncetohear,waveformer,clarityChallenge}, but these are evaluated on devices with much higher clock frequencies, power budgets, and memory footprints than our target hardware. The most relevant, FSPEN~\cite{fspen}, enhances speech in real-time using gated recurrent units (GRUs). Our results show that replacing its bidirectional GRU with an all-MLP layer improves efficiency on our target hardware platform.


 
\vskip 0.02in\noindent {\bf TinyML.} In the audio domain, prior work has applied TinyML methods to classification tasks like keyword spotting~\cite{kws}, speaker verification~\cite{tinysv}, and sound event detection~\cite{bose}, as well as regression tasks like speech enhancement and denoising~\cite{tinylstms, tinydenoiser,fqse,TRunet,10248154,10446796,8489456}. These networks however are not designed for speech separation.  They also do not process the time and frequency components as individual sequences as has become an essential component in state-of-the-art speech separation models. \cite{fqss} designs a quantized audio separation network but the proposed network is neither causal, real-time nor can run on low-power accelerators.  



\section{Methods}



%

\subsection{System Requirements and Runtime Decomposition}



Real-time on-device enhanced hearing applications impose strict constraints on model size, runtime, and power consumption. For example, if our model receives 6~ms long audio chunks, running inference on these chunks should take less than 6~ms for real-time operation. Further, non-volatile storage is limited, e.g. given the size of GAP9 eMRAM, the model size can be at most 1.5~MB to avoid using additional memory components. Finally, to ensure $\gt$ 6 hours of continuous use on a 675 hearing aid battery,  power consumption  must stay below  100~mW.



\begin{figure}[t!]
\centering
\includegraphics[width=\linewidth]{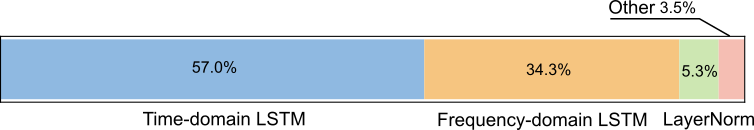}
\vskip -0.11in
\caption{{Runtime profile of an existing dual-path model.}}
\label{fig:runtime}
\vskip -0.24in
\end{figure}

We start with a causal dual-path model, TF-GridNet, that operates on the TF-domain~\cite{tfgridnet}. {Given our target hardware, we remove the self-attention module but keep the recurrent modules. We quantize all nodes in TF-Gridnet into \textsc{int8}, with the exception of Layer Normalization and the batched time-domain LSTM, which are quantized to \textsc{float16} since these layers lack \textsc{int8} out-of-the-box support.} Fig.~\ref{fig:runtime} profiles the runtime on GAP9 running at 370~MHz. This model requires 23.5~ms to process one 6~ms chunk. The major contributors to the runtime are 1) the frequency-domain bidirectional LSTMs and 2) the time-domain batched unidirectional LSTMs.

\subsection{TF-MLPNet}

TF-MLPNet is a tiny real-time  network for speech separation. Our network has two main  components:  a conv-batched LSTM, that enables parallel inference of a batch of inputs to an LSTM using convolutional layers, and  a highly parallel all-MLP-Mixer module that replaces the sequential bidirectional LSTM. 

In TF-MLPNet, we first apply a short-time Fourier transform (STFT) to convert a time-domain audio signal $x \in \mathbf{R}^{1 \times t}$ into a time-frequency (TF) representation $X'\in\mathbf{C}^{1 \times F \times T}$ over $T$ frames and $F$ frequency bins. The real and imaginary components are concatenated along the channel dimension to get $X \in \mathbf{R}^{2 \times F \times T}$. As shown in Fig.~\ref{fig:architecture}(a), a causal   $3 \times 3$  2D convolution encoder extracts   a $C$-dimensional latent representation $\hat{X} \in \mathbf{R}^{C \times F' \times T'}$.  $\hat{X}$ is   processed by $B$ MLPNet blocks, each containing an MLP-Mixer and a conv-batched LSTM module, producing $\Tilde{X} \in \mathbf{R}^{1 \times F' \times T'}$. A causal $3 \times 3$ 2D transposed convolution decoder then recovers the TF representations $Y \in \mathbf{R}^{2S \times F \times T}$ for each of the $S$ speakers ($S=2$ for BSS and $S=1$ for TSE). This produces two channels per speaker — the first half corresponds to the real components of the TF representation, while the second half corresponds to the imaginary components. Accordingly, we reinterpret $Y$ as $Y' \in\mathbf{C}^{S \times F \times T} $ use inverse STFT and overlap-add operations to reconstruct the time-domain output signal $y \in \mathbf{R}^{1 \times t}$ for each speaker. 

\subsubsection{MLP-Mixer Module}

At the spectral stage of existing  dual-path models~\cite{tfgridnet}, the key  contributor to  the inference time is the bidirectional frequency-domain LSTM. 
Due to the nature of RNNs, each frequency bin in the input must be processed sequentially. One way to reduce the runtime is to compress the number of frequencies with strided convolutions before and after frequency-domain processing~\cite{soundbubble}. However, this was shown to reduce the algorithm's  performance. An alternative approach is to replace the recurrent architecture with one that can better utilize the parallel processing capabilities of the neural accelerator. While this can be achieved using transformers or linear RNNs, today's low-power hardware accelerators do not support these complex operations.

We instead replace the bidirectional frequency-domain LSTM with an MLP-Mixer~\cite{mlpmixer}, that applies a sequence of multilayer perceptions alternately along the frequency and channel dimensions (Fig.~\ref{fig:architecture}). This MLPMixer module is repeated $M$ times successively in every MLPNet Block. To further reduce inference time, we replace the GELU nonlinearities with simpler ReLU nonlinearities and omit the layer normalization. 



\subsubsection{Conv-Batched LSTM }

Existing dual-path models~\cite{tfgridnet} process frequency components independently in the time domain. During training, GPUs handle this efficiently by treating each frequency bins as a  separate batch. However, inference on low-power accelerators is challenging since many lack support for batched processing.


During streaming inference, the batched LSTM receives $\hat{X} \in \mathbf{R^{C \times F' \times 1}}$ and processes a single LSTM step on $F'$  independent sequences. A naive solution is sequential processing per frequency bin to maintain a batch size of 1, but this underutilizes parallel processing and greatly increases inference time.


Instead of a standard LSTM kernel, we decompose the LSTM into its building blocks. We use 1D convolutions (kernel size = 1) for the linear gates, treating the frequency dimension as the sequence dimension. This allows parallel processing of all frequencies without special hardware support.  Note that since we process only a single time frame at a time during streaming inference, batched inference for frequency-domain processing is not required, and so the batch size, i.e., the number of independent time frames, is just 1.

\subsection{Mixed-Precision Quantization}\label{sec:quant}

We use Quantization-Aware Training (QAT)~\cite{gholami2022survey} to simulate quantization errors during training and reduce performance degradation. We start with a floating-point model and fine-tune it using the FQSE quantization framework~\cite{fqse}. Our weight quantization is symmetric and per-channel, while activation quantization is asymmetric and per-tensor.

Our fully \textsc{int8}-quantized network with QAT produces a noticeable performance degradation. To bridge the gap between quantized and floating-point models, we designed a mixed-precision QAT approach to balance performance and runtime. 

We quantize the first input convolution and the last deconvolution layers to \textsc{bfloat16} instead of \textsc{int8} to preserve the high-precision information from the input in the first layer and to reconstruct high-quality audio with the last deconvolution. Additionally, our experiments with batched LSTMs revealed that convolution layers dominate the runtime and that quantization errors accumulate temporally. {So, we use a mixed-precision LSTM, where we quantize convolution layers to \textsc{int8} for efficiency while keeping activation, addition, multiplication operations, and cell states in \textsc{bfloat16} to minimize quantization noise and improve performance. }

To ensure real-time operation while minimizing performance loss, we use \textsc{int16} activations for the MLP modules at odd-numbered MLPNet blocks and \textsc{int8} activations for those at even-numbered blocks. The MLP module weights are always quantized in \textsc{int8}. Finally, we incorporate the SDR-aware knowledge distillation loss function~\cite{fqss} into the QAT process.

\begin{table}[t!]
  \caption{Main results. SISDR, PESQ and DNS-MOS are reported for \textsc{fp32} BSS models. For runtime, we assume the networks are fully \textsc{int8} quantized, except in the case where they  have Layer Normalization, which is quantized to \textsc{fp16}. Values in red do not meet our system requirements.}
  \label{tab:main-res}
  \vskip -0.1in
  \centering
\setlength{\tabcolsep}{2.1pt}
 
  \begin{tabular}{ c c c c c c }
    \toprule
    \multicolumn{1}{c}{\textbf{Name}} &
    \multicolumn{1}{c}{\textbf{Runtime}} & 
    \multicolumn{1}{c}{\textbf{SISDR}} &
    \multicolumn{1}{c}{\textbf{PESQ}} &
    \multicolumn{1}{c}{\textbf{DNS}} & 
    \multicolumn{1}{c}{\textbf{Param}} \\
    
    & 
    \textbf{(ms)}&
    \textbf{(dB)} &
    &
    \textbf{-MOS}&
    \textbf{(K)}\\
    \midrule
    Mixture & -- & 0.00 & 1.24 & 2.48 & -- \\
    \midrule
    TF-GridNet~\cite{tfgridnet} & \textcolor{darkred}{16.4} & 14.78  & 2.39 & 3.28 & 173 \\
    TFG-LN & \textcolor{darkred}{15.1} & 14.08 & 2.29 & 3.12 & 173  \\
    TFG-LN+2F  & \textcolor{darkred}{8.8} & 13.78 & 2.27 & 3.14 & 198 \\ 
    TFG-LN+4F & 5.6 & 13.51 & \textbf{2.25} & 3.09 & 222  \\
    TFG-LN+6F & 4.5 & 13.16 & 2.17 & 3.07 & 247  \\
    \midrule
    TFG-LN+GRU & \textcolor{darkred}{12.4}  & 14.47 & 2.35 & 3.22 & 147  \\
    TFG-LN+2F+GRU  & \textcolor{darkred}{7.4} & 12.87 & 2.13 & 3.00 & 172 \\
    TFG-LN+4F+GRU & 4.9  & 12.66 & 2.10 & 2.96 & 197 \\
    \midrule
    TF-MLPNet & 3.6 & \textbf{14.12} & 2.23 & \textbf{3.21} & 493 \\
    TF-MLPNet+2F  & 2.8  & 13.06 & 2.09 & 3.04 & 215\\
    \bottomrule
  \end{tabular}
    \vskip -0.2in
\end{table}

\section{Experiments and results}



\vskip 0.03in\noindent{\bf Datasets.} We train our model using mixtures from LibriSpeech~\cite{librispeech} and evaluate it on the LibriSpeech test set~\cite{librispeech} and VCTK~\cite{vctk}. Each 5-second, 16 kHz speech mixture is created by sampling two different speaker utterances from the same corpus split. Utterances longer than 5 seconds are cropped, while shorter ones are padded with silence. For TSE, one speaker is chosen as the target, and the speaker d-vector~\cite{dvector} embedding is computed from a different utterance by that target speaker. The interfering speech is scaled to achieve an input SNR uniformly distributed in [-10,10] dB. Training speech files come from \verb|train-clean-360|, validation from \verb|dev-clean|, and testing from \verb|test-clean|. The training set is generated on-the-fly, while validation and test sets contain 2k and 1k mixtures, respectively.

\vskip 0.03in\noindent{\bf Evaluation setup.} We compare with multiple model variants.  For TF-GridNet, we use the causal implementation from~\cite{lookoncetohear} without self-attention and with  hyperparameters $B=6$, $D=32$ and $H=32$. We remove the LayerNormalization modules to obtain the model TFG-LN. We further introduced frequency compression on the frequency-domain processing component used in~\cite{soundbubble}, and we refer to the resulting model with a frequency compression rate $\alpha$ as TFG-LN+$\alpha$F.  We also considered a variant where we replaced the LSTM with a GRU, referred to as TFG-LN+$\alpha$F+GRU. In addition to the hyperparameters enumerated above, the TF-MLPNet used in our experiments has an MLP-Mixer with $M=2$ MLP-Mixer repetitions. Additionally, when we apply frequency compression to our TF-MLPNet architecture, we refer to it as TF-MLPNet+$\alpha$F. 

Following~\cite{lowlatse}, we use a 10 ms output window and a 6 ms hop size, resulting in a 10 ms algorithmic latency. For BSS, the decoder outputs two channels (one per speaker), while TSE uses a single channel. In TSE, the model is conditioned on d-vector embeddings via a FiLM layer after the encoder. We convert PyTorch models into optimized kernels for GAP9 using GAPFlow.

\vskip 0.03in\noindent{\bf Loss function and training hyper-parameters.} We train BSS models using Permutation Invariant Training with negative SI-SDR as the loss function. For TSE, we use a combined loss: $L_{SI-SDR} + L_{PESQ}$, where $L_{SI-SDR}$ is   negative SI-SDR, and $L_{{PESQ}}$ is calculated using \verb|torch-pesq|.  Each epoch processes 20k mixtures before validation. Models are trained for 400 epochs using a three-stage schedule: (1) linearly increasing the learning rate from 1e-4 to 1e-3 over 10 epochs, (2) maintaining 1e-3 for 200 epochs, and (3) halving it every 30 epochs for the remaining 190. Model parameters are optimized using AdamW and we use a gradient clipping of 0.1. Model performance is evaluated using the  the epoch with the lowest validation loss.





\vskip 0.03in\noindent{\bf QAT hyper-parameters.}  We start with the trained floating-point parameters and fine-tune it for 100 epochs. Since QAT is time-consuming, each epoch processes 4k mixtures. Training starts with a 1e-3 learning rate using a \verb|ReduceLROnPlateau| scheduler (patience = 5, factor = 0.5). Early stopping is triggered if validation loss does not improve for 20 epochs.

\begin{table}[t!]
  \caption{ SI-SDRi (dB) for blind source separation (BSS) as a function of percentage of training set used for training models.}
  \label{tab:datasize}
  \vskip -0.1in
  \centering
\setlength{\tabcolsep}{2.5pt}
 
  \begin{tabular}{ c c c c c c c c c }
    \toprule
     & \multicolumn{7}{c}{\textbf{Training Dataset Percentage (\%)}}
     \\
    \midrule
    \multicolumn{1}{c}{\textbf{Model}} & 
    \multicolumn{1}{c}{\textbf{1}} &
    \multicolumn{1}{c}{\textbf{2}} &
    \multicolumn{1}{c}{\textbf{5}} & 
    \multicolumn{1}{c}{\textbf{10}} & 
    \multicolumn{1}{c}{\textbf{25}} & 
    \multicolumn{1}{c}{\textbf{50}} & 
    \multicolumn{1}{c}{\textbf{100}} & \\
    
    \midrule
    TFG-LN+4F & 4.52 & 7.83 & 11.44 & 12.99 & 13.34 & 13.26 & 13.51 \\
    TF-MLPNet & 4.18 & 6.08 & 9.65 & 12.99 & 13.85 & 14.04 & 14.12 \\
    \bottomrule
  \end{tabular}
    \vskip -0.2in
\end{table}


\vskip 0.05in\noindent{\bf Results.} Table~\ref{tab:main-res} compares floating-point performance  of different models on the BSS task  using SI-SDR, PESQ, and DNSMOS OVRL~\cite{dnsmos}. We measure quantized runtime for processing a 6 ms audio chunk on GAP9 at 370 MHz.


The original TF-GridNet achieves the best floating-point performance (14.78 dB) but requires 25.8 ms runtime at \textsc{fp16}, reduced to 16.4 ms with \textsc{int8} quantization and \textsc{fp16} layer normalization. Removing LayerNorm (TFG-LN) and applying our Conv-Batched LSTM method  cuts \textsc{fp16} runtime to 22.3 ms, with full \textsc{int8} quantization reducing it to 15.1 ms—still too slow for real-time use. Frequency compression  meets real-time requirements, bringing \textsc{int8} runtime down to 5.6 ms ($\times 4$ compression) and 4.5 ms ($\times 6$ compression), but at a 0.5-1 dB floating-point SI-SDR performance drop. Replacing LSTM with GRU also speeds up \textsc{int8} runtime but affects performance.


Our TF-MLPNet model, which replaces frequency-domain LSTMs with MLP-Mixer, achieves a drastic runtime reduction to 3.6 ms with \textsc{int8} quantization. This improvement is due to its use of simple, parallelizable MLPs that efficiently leverage the neural accelerator. TF-MLPNet offers the lowest runtime while maintaining a floating-point performance at 14.12 dB. {A paired t-test was conducted between our TF-MLPNet and other baselines which meets the real-time requirements for each metric, showing a significant difference with $p < 0.05$.}

We analyze how training dataset size affects TF-MLPNet performance compared to TF-LN+4F, the best real-time baseline model variant. In Table~\ref{tab:datasize}, both models are trained for the same number of epochs, but with varying proportions of speakers seen during training. While TF-MLPNet underperforms TF-LN+4F with limited data, it surpasses it as more speakers are included in training. This indicates that TF-MLPNet scales better with increased data, which matches the observation in~\cite{mlpmixer}.

Quantization impacts model performance, so we experimented with different quantization settings and QAT techniques to assess their effects on performance and runtime.  We evaluate  the following configurations: (1) FP32: original float-pointing model, (2) INT8: fully quantized INT8 model, (3) MixLSTM: Mixed-precision LSTM (see~\xref{sec:quant}), with other modules quantized to INT8, (4) MixLSTM+FPConv: mixed-precision LSTM, BFLOAT16 Conv/Deconv, other modules quantized to INT8, (5) MixLSTM+FPConv+MixMLP: mixed-precision LSTM, BFLOAT16 Conv/Deconv, mixed-precision MLP (see~\xref{sec:quant}), others in INT8, and (6) MixLSTM+FPConv+FULLMLP: mixed-precision LSTM, BFLOAT16 Conv/Deconv, fully INT16-quantized MLP, others in INT8. We trained with SI-SDR loss and also explored SDR-aware knowledge distillation ("+KD"). As shown in Table~\ref{tab:main-qat}, full INT8 quantization with standard QAT led to around 4 dB drop in SI-SDRi. However, incorporating KD loss and mixed-precision quantization—including mixed-precision LSTM, BFLOAT16 Conv, and mixed-precision MLP—recovered SI-SDRi to 13.52 dB while still achieving real-time performance.


\begin{table}[t!]
  \caption{Quantization results. We measure the SISDRi, model size, runtime and power consumption for different QAT strategies. MixLSTM refers to quantizing convolution layers to \textsc{int8} while keeping others in \textsc{bfloat16}. FPConv refers to quantizing the input convolution and output deconvolution in \textsc{bfloat16} instead of \textsc{int8}. KD refers to using SDR-aware knowledge distillation as the training loss. MixMLP refers to quantizing only half the MLP modules into INT16, while in FullMLP, we quantize all of them to INT16. The reported model size includes the memory needed to store the quantized model parameters, but not the quantization constants.}
  \label{tab:main-qat}
  \vskip -0.1in
  \centering
\setlength{\tabcolsep}{2.5pt}
 
  \begin{tabular}{ l c c c c }
    \toprule
    \multicolumn{1}{l}{\textbf{QAT config}} & 
    \multicolumn{1}{c}{\textbf{SISDRi}} &
    \multicolumn{1}{c}{\textbf{Size}} &
    \multicolumn{1}{c}{\textbf{Runtime}}&
    \multicolumn{1}{c}{\textbf{Power}}\\
    \textbf{TF-MLPNet}& 
    \textbf{(dB)}&
    \textbf{(kB)}&
    \textbf{(ms)}&
    \textbf{(mW)}\\
    \midrule
    FP32 & 14.12 & 1926 & -- & --\\
    INT8 & 10.21 & 481 & 3.6 & 54.9\\ 
    \quad +MixLSTM & 11.22 & 481 & 4.0 & 58.1 \\ 
    \quad \quad +FPConv &  12.57 & 483  & 4.2 & 60.7 \\ 
    \quad \quad \quad +KD  & 13.07 & 483 & 4.2 & 60.7 \\ 
    \quad \quad \quad \quad +MixMLP   & 13.52 & 483  & 5.6 & 80.1 \\ 
    \quad \quad \quad \quad +FullMLP & 13.65  & 483 & \textcolor{darkred}{6.5} & \textcolor{darkred}{--} \\
    \bottomrule
  \end{tabular}
    \vskip -0.2in
\end{table}

Table~\ref{tab:main-vctk} shows that TF-MLPNet's performance gains generalize to out-of-distribution datasets. We created 5-second audio mixtures using VCTK data, following the same process as before. BSS models trained on LibriSpeech mixtures were then evaluated on the VCTK mixtures. TF-MLPNet  outperformed baseline models, showing   generalization across datasets.

\begin{table}[t!]
  \caption{BSS results on 2-speaker mixtures from VCTK. All models were only trained on data from LibriSpeech.  SISDR, PESQ and DNS-MOS are reported for \textsc{fp32} BSS models. For runtime, we assume the networks are fully INT8 quantized.}
  \label{tab:main-vctk}
  \vskip -0.1in
  \centering
\setlength{\tabcolsep}{5pt}
 
  \begin{tabular}{ c c c c }
    \toprule
    \multicolumn{1}{c}{\textbf{Name}} & 
    \multicolumn{1}{c}{\textbf{SISDR}} &
    \multicolumn{1}{c}{\textbf{PESQ}} &
    \multicolumn{1}{c}{\textbf{Runtime}}\\
    
    & 
    \textbf{(dB)}&
    & \textbf{(ms)}\\
    \midrule
    Mixture & -0.01 & 1.55 & -- \\
    \midrule
    TFG-LN+4F & 12.36 & \textbf{1.97} & 5.6 \\
    TFG-LN+6F & 11.51 & 1.94 & 4.5 \\
    TFG-LN+4F+GRU & 11.68 & 1.89 & 4.9\\
    \midrule
    TF-MLPNet & \textbf{12.63} & 1.95 & 3.6\\
    TF-MLPNet+2F & 11.59 & 1.85 & 2.8 \\
    \bottomrule
  \end{tabular}
\end{table}

Finally, we evaluated TF-MLPNet on the TSE task, comparing it against baseline models and prior work, including pDCCRN~\cite{pdccrn} and TinyDenoiser~\cite{tinydenoiser}. For a fair comparison, all models used the same STFT configuration and a FiLM layer right after the encoder for target speaker conditioning. pDCCRN had the number of convolution filters set to  $[16, 32, 64, 128, 256, 256]$, a $5\times2$ kernel with a $2\times1$ stride, and an LSTM hidden size of 256. For TinyDenoiser, aside from the STFT configuration, we use the same hyperparameters described in~\cite{tinydenoiser}. TF-MLPNet outperformed all models across metrics while maintaining real-time performance, demonstrating its effectiveness across both speech separation tasks.


\begin{table}[t!]
  \caption{Target Speech extraction (TSE) results.  SISDR, PESQ and DNS-MOS are reported for \textsc{fp32} TSE models. For runtime, we assume the networks are fully INT8 quantized. Values in red do not meet our runtime or memory requirements.}
  \label{tab:tse}
  \vskip -0.1in
  \centering
\setlength{\tabcolsep}{2.4pt}
 
  \begin{tabular}{ c c c c c c }
    \toprule
    \multicolumn{1}{c}{\textbf{Name}} & 
    \multicolumn{1}{c}{\textbf{SISDR}} &
    \multicolumn{1}{c}{\textbf{PESQ}} &
    \multicolumn{1}{c}{\textbf{DNS}} & 
    \multicolumn{1}{c}{\textbf{Param}} &
    \multicolumn{1}{c}{\textbf{Runtime}}\\
    
    & 
    \textbf{(dB)}&
    &
    \textbf{-MOS}&
    \textbf{(K)}&
    \textbf{(ms)}\\
    \midrule
    Mixture & 0.05 & 1.26 & 2.50 & -- & --\\
    \midrule
    pDCCRN~\cite{pdccrn} & 10.71 & 2.15 & 3.05 & \textcolor{darkred}{3218} & \textcolor{darkred}{--}
    \\ 
    \midrule
    TFG-LN+2F & 12.22 & 2.32 & 3.23 & 213 & \textcolor{darkred}{8.9} \\ 
    TFG-LN+4F & 11.90 & 2.28 & 3.18 & 238 & 5.7 \\
    TFG-LN+6F & 11.28 & 2.24 & 3.15 & 263 & 4.6 \\
    TFG-LN+2F+GRU & 12.20 & 2.32 & 3.18 & 188 & \textcolor{darkred}{7.5} \\ 
    TFG-LN+4F+GRU & 12.00 & 2.30 & 3.15 & 213 & 5.0 \\ 
    \midrule
    TinyDenoiser~\cite{tinydenoiser} & 8.15 & 1.70
    & 2.65 & 1056 & 0.464 \\ 
    \midrule
    TF-MLPNet & \textbf{12.37} & \textbf{2.37} & \textbf{3.32} & 509 & 3.6 \\
    TF-MLPNet+2F & 11.77 & 2.18 & 3.12 & 231 & 2.9 \\
    \bottomrule
  \end{tabular}
    \vskip -0.2in
\end{table}

\section{Conclusion}

We introduce TF-MLPNet, the first real-time speech separation network for low-power hearables, outperforming existing streaming architectures. While we use hardware consistent with prior research~\cite{neuralaids}, exploring our methods on platforms like Qualcomm's S7 series, Analog Devices' MAX78002, and Syntiant's NDP120 offers interesting future directions. Further work includes enabling other audio tasks such as target sound extraction, distance-based multi-channel source separation~\cite{soundbubble}, and directional hearing~\cite{directional} on constrained hardware.

\noindent{\bf Acknowledgments.} The researchers are partly supported
by the Moore Inventor Fellow award \#10617, Thomas J. Cable Endowed Professorship, and a UW CoMotion innovation gap fund. This work was facilitated through the use of computational, storage, and networking infrastructure provided by the UW HYAK Consortium.



\bibliographystyle{IEEEbib-abbrev}
{\footnotesize
\bibliography{refs}
}
\end{document}